\documentstyle{mn}

\title[The Near-Infrared Fundamental Plane of Ellipticals]
{Near-Infrared Fundamental Plane of Elliptical Galaxies}

\author[B. Mobasher et al.]
{B.~Mobasher$^1$, R.~Guzm\'an$^2$, A.~Arag\'on-Salamanca$^3$, S.~Zepf$^4$\\
$^1$Astrophysics Group, Blackett Laboratory, Imperial College, Prince Consort Rd, London SW7 2BZ, UK\\
$^2$UCO/Lick Observatory, University of California, Santa Cruz, 1156 High Street, Santa Cruz, CA 95064, USA\\
$^3$Institute of Astronomy, Madingley Road, Cambridge CB3 0HA, UK\\
$^4$Department of Astronomy, University of California, Berkeley, CA 94720, USA\\}

\begin{document}
 
\maketitle
\begin{abstract}
Near-infrared (2.2 $\mu m$) observations of a sample of 48 elliptical
galaxies in the Coma cluster have been carried out and used to study
the near-ir fundamental plane (FP) of ellipticals in this cluster. 
An rms scatter
of 0.072 dex was found for this relation, similar to that of its 
optical counterpart, using the same sample of galaxies. This corresponds to
an uncertainty of $18\%$ in distances to individual galaxies derived
from this relation. 
The sensitivity of the near-infrared FP to the star formation or changes in 
metallicity and stellar population among the ellipticals was explored and 
found to be small. Although, a likely source of scatter in this relation 
is contributions from the
Asymptotic Giant Branch (AGB) population to their near-infrared light. 
Allowing for observational uncertainties, we find an intrinsic 
scatter of 0.060 dex in the near-ir FP. The cluster galaxies presented here, 
provide the zero-point for the peculiar velocity studies, using the 
near-infrared FP. 

Changes in the slopes of the $D-\sigma$ and $L-\sigma$ relations of ellipticals
between the optical and near-infrared wavelengths were investigated and 
found to be due to variations in metallicity or age 
(or a combination of them). 
However, it was not possible to disentangle the effects of age and metallicity 
in these relations. 

We find $M/L\propto M^{\alpha}$ with $\alpha=0.18\pm 0.01$ at the near-ir and
$\alpha=0.23\pm 0.01$ at the optical wavelengths, using the same sample
of galaxies. This relation is interpreted as 
due to a mass-metallicity effect or changes in age or the IMF slope 
with mass. Using evolutionary population synthesis models, we find that
the effects of age and metallicity decouple on the $(M/L)_K$ vs. Mg$_2$ and
$(M/L)_K$ vs. (V$-$K) diagrams. The models suggest that the observed trends 
on these relations may  be due to an age sequence while metallicity mainly 
contributing to the scatter. 
\end{abstract}
 
 
\section{Introduction} 
Elliptical galaxies define a Fundamental Plane (FP) on the 3-parameter space
of the half-light radius $R_e$, the mean surface brightness within that
radius $\langle SB \rangle_e$, and the central velocity dispersion $\sigma$ (Djorgovski
\& Davis 1987; Dressler et al. 1987a). The scatter around the FP is only 
$\sim0.08$ dex in log$R_e$ for cluster ellipticals.
The existence of this scaling relation provides one of the most important 
constraints to model the formation and evolution of elliptical galaxies 
(Bender et al 1993; Guzm\'an et al. 1993; Renzini \& Ciotti 1993; 
Zepf \& Silk 1996). 
Also, it can be used as a distance indicator to 
trace the non-Hubble motions in the local universe (Dressler et al 1987b; 
Lucey \& Carter 1988; Lynden-Bell et al. 1988), 
and potentially as a cosmological evolutionary probe when studying 
the FP for cluster ellipticals at higher redshifts (Franx et al. 1996; 
van Dokkum \& Franx 1996; Barger et al. 1996). 

	The biparametric nature of elliptical galaxies suggests that the virial
theorem is the main constraint on their structure (Faber et al. 1987; 
Djorgovski 1987; but see Guzm\'an et al. 1993). The tilt of the FP relative
to the virial equation implies a systematic variation either in the IMF, the
shape of the light profile, the age and metallicity or the dark and luminous 
matter distribution with
galaxy mass (Guzm\'an et al. 1993; Ciotti et al. 1996). Our understanding
of the physical mechanisms responsible for such variation is still at a very 
early stage. Moreover, substantial fine-tuning is required to produce
the tilt while preserving the small scatter of the galaxy distribution along
the FP (Ciotti et al. 1996).  
The tightness of the FP for cluster ellipticals is clear evidence
for a very standardized and synchronized production of ellipticals, with the
vast majority of their stellar populations being formed at redshifts $z>2$, 
at least for those ellipticals in clusters (Renzini 1995; Ellis et al. 1996). 
However, there is 
growing evidence that ellipticals outside the core of rich clusters tend to 
have a younger stellar component (Rose 1985; Pickles
1985; Bower et al. 1990). As a result, non-cluster ellipticals 
at a given $R_e$ and/or $\sigma$ have, on
average, bluer colours, higher effective surface brightnesses, and
lower $Mg_2$ line strength indices as compared to their cluster counterparts
(Larson et al. 1980; Guzm\'an et al. 1992; de Carvalho \& Djorgovski 1992).
This systematic variation of stellar population properties due to the 
environment may in turn translate into a zero-point offset of the FP-based
distance indicators (such as $D_n-\sigma$) and thus lead to spurious peculiar
velocities (Kaiser 1988; Silk 1989; de Carvalho \& Djorgovski 1992; but see 
Burstein et al. 1990). Indeed, using stellar population evolutionary models, 
Guzm\'an \& Lucey (1993) have combined the FP with the $\sigma-Mg_2$ relation 
to create a new distance indicator that is independent of age/environmental 
differences. When the new distance indicator is applied to the ellipticals
with a younger stellar component, the large positive peculiar motions 
previously reported are greatly removed (Guzm\'an \& Lucey 1993). 

	Most studies on the FP have been done at optical wavelengths. 
Observations in the infrared may shed light on the origin of the tilt and
scatter of the FP by probing a wavelength region with a different sensitivity 
to stellar population effects. In particular, the effect of age and/or
metallicity variations in ellipticals is expected to be greatly reduced 
in the K-band 
(i.e., 2.2$\mu m$) since the galaxy light at these wavelengths is mainly 
sensitive to the old stellar component and is less affected by line-blanketing.
Moreover, the K-band is significantly less sensitive to both the presence of 
dust in ellipticals and galactic extinction. The new K-band FP can also be used
as an independent distance indicator to check for stellar-population effects
on the peculiar velocity measurements derived using the FP in the optical. 
Early work on the infrared FP 
relied on single element photometry and hence, lacked detailed surface 
photometry (Gregg 1995a; Recillas-Cruz et al. 1990, 1991). Recently, Pahre et 
al. (1996) have shown that the coefficients and scatter of the FP in the 
K-band for a sample of ellipticals in five nearby clusters are similar to 
those derived in the optical. They conclude that the observed small 
differences are consistent with the reduction of metallicity effects in the 
near-infared bandpass.  

In this paper, we present the first results of a large-scale K-band survey
of elliptical galaxies located in different environments (i.e., clusters, field
and Great Attractor region). Here we study the 
coefficients and scatter of the K-band FP, using accurate surface
photometry for a homogeneous sample of 48 galaxies in the Coma cluster. 
By considering a single cluster, we avoid introducing artificial scatter 
due to distance errors or non-universality of the FP.
The aim of this study is three fold: 1) to shed light on the origin of the FP
by using a wavelength region with a different sensitivity to the stellar
population effects than the optical; 2) to establish the intrinsic scatter
on the K-band FP; and 3) to set the zero-point of the new infrared distance 
scale using the K-band FP for Coma cluster ellipticals.

In the next section, the observations and data reduction are discussed. This
is followed by the study of the near-infrared scaling laws in section 3.
The implications for the origin of the FP and the new distance scale are
discussed in section 4. Finally, our conclusions are summarized in section 5.

\section{Observations and Data Reduction}

\subsection {Observations}

The observations used in this study were made at the United Kingdom Infrared
Telescope (UKIRT) in the period 28-30th April 1994. Near-infrared K-band (2.2 
$\mu m$) images of 48 elliptical galaxies in the Coma cluster were obtained, using the
$256 \times 256$ infrared array detector (IRCAM3) with a pixel size of 
$0.286''$. The galaxies are selected to be ellipticals (i.e. no lenticulars) 
from the sample in Lucey et al (1991, Table 6) with reliable 
velocity dispersion and optical photometry. They are located at different 
distances from the center of the Coma cluster ($\alpha=$ 12 57 19  
$\delta=$ 28 15 51 (1950)). 
Also, they have optical half-light diameters smaller than $1'$,
fitting the field of view of the IRCAM3. An exposure time of 10 sec. was
used with  6 co-adds. For each object, nine exposures were taken at
different positions, separated by $20''$,   
giving a total integration time of 9 mins. per object. 
This technique was employed to avoid the
effect of bad pixels and cosmic rays, to construct reliable flatfields for
each frame and to measure the sky background. 
Surface photometry of smaller galaxies can be affected by the seeing 
condition which was estimated to be around $1''$.
The sensitivity of the results to this effect will be explored in the 
following analysis. 
Observations of faint UKIRT standards were also carried out during the night
and were used to monitor the accuracy of the photometry. 

\subsection {Data Reductions.}

Observations were dark subtracted using the dark frames closest (in time) to
each object. For each galaxy, the nine mosaic frames were then median filtered 
and the result normalised to its median pixel value to construct the 
flatfield. The dark subtracted, flatfielded frames were 
then registered and 
mosaiced to increase the S/N. The data reduction 
was carried out, using the CCDPACK software in the STARLINK environment. 
The median sky value was then
estimated for each frame individually and used to sky-subtract that frame.
The mosaic frames, providing a larger sky coverage than single frames, 
were used for this purpose and the sky values were 
measured as far away from the center of galaxy as possible. The variations
in sky estimates at different positions on the frame were monitored, 
giving a measure of the uncertainties in these values. This did not exceed
0.02 mag., even for the bigger galaxies for which the uncertainties are
larger. Therefore, this is taken as our estimated error due to sky
subtraction. 

The standard stars were reduced in the same way and used to establish the
extinction relation for each night. Changes in the zero-points, estimated from
the standards throughout the run (internal photometric error), is 
0.02 mag.   

\subsection {Photometry}

Aperture photometry was performed on the final images and used to
construct the near-ir growth curves for individual galaxies. The magnitudes
were corrected for atmospheric extinction, Galactic absorption (taking
$A_K=0.085 A_B$ (Pahre et al. 1996) 
and the median $A_B$ for the Coma galaxies from Faber et al
(1989)) and redshift effect (assuming $k(z)= 3.3z$ for ellipticals at
the K-band (Persson et al (1979)). 
The photometric accuracy is checked by comparing magnitudes 
of the objects in our sample with the same galaxies from other independent
observations. The K-band magnitudes
for 18 galaxies in the present sample, in common with Bower et al (1992),
are measured over an aperture of $17''$ diameter, used by these authors.
The two magnitudes are compared in Figure 1a and show a mean offset of
$$\langle K_{this\ study} - K_{Bower\ et\ al.}\rangle = -0.06 \pm 0.01 $$
with an rms scatter of 0.03 mag. Assuming equal errors in both the data sets, 
this corresponds to an internal error of $0.03/\sqrt 2 = 0.02$ mag. per
galaxy, consistent with that claimed by Bower et al (1992). 
The cause of the zero-point offset between the two data sets is not clear
but is likely to be due to different detectors used. However, since the data
here are 2-D and Bower et al's is based on single element photometry, 
we expect to have a more accurate control over the sky subtraction in 
our data. Neverthless, the offset is relatively small. 

The growth curves, corrected for Galactic absorption and redshift effects, were
used to construct near-infrared surface brightness
profiles for each galaxy, corrected for the $(1+z)^{-4}$ surface brightness 
dimming. Near-infrared isophotal diameters, corresponding to 
a mean surface brightness of 16.5 mag/arcsec$^2$  
are then estimated. This isophote was adopted 
to be directly comparable with   
the mean optical surface brightness of 19.8
mag/arcsec.$^2$ (assuming $V-K=3.3$ mag. for ellipticals) used in  
the optical $D_V-\sigma$ relation. 
To examine the accuracy of our surface photometry, we have also 
obtained data for a sub-set of 12 ellipticals in our sample, using
the IRIS at the Anglo-Australian Telescope (March 1995). These diameters, 
measured at a mean K-band surface brightness 
of 16.5 mag./arcsec$^2$, are compared with the present measurements
in Figure 1b and show a mean offset of
$$\langle  log (D_K^{UKIRT}/D_K^{AAT}) \rangle = 0.013\ dex \pm 0.006 dex $$
with an rms scatter of 0.019 dex. Assuming that the uncertainty in these measurements
is shared equally between both the data sets, we estimate an observational error
of 0.01 dex in log($D_K$) values. 

The diameters of smaller galaxies in our sample might be affected by
the seeing condition. To estimate this effect, we used the models in
Lucey et al (1991) and estimated the effect of 1 arcsec seeing on the 
aperture photometry (the observations here were done in 0.8-1.2 arcsec seeing
conditions). The correction was then applied to one of our smaller
galaxies and its photometric parameters were recalculated. We find that 
at 1 arcsec
seeing conditions, an increase in log($D_K$) of only 0.01 dex was needed 
due to seeing. The
diameters of larger galaxies were almost insensitive to this effect. 
The range in  
surface brightness values between 16 and 17 mag/arcsec$^2$ was explored in 0.1 
intervals. The results in the following sections were found to be insensitive 
to the actual choice of the isophote. 

The near-infrared effective diameter and surface brightness (i.e. the diameter containing
half the total luminosity of a galaxy and the mean surface brightness within that
diameter) were calculated for galaxies in the present sample by fitting the
observed K-band curve of growths to the de Vaucouleurs $r^{1/4}$ law. 
The accuracy of this fitting procedure was estimated by generating
simulated profiles following the $r^{1/4}$ law with known effective parameters 
and artificially added noise, analogous to the photometric errors. 
An $r^{1/4}$ fit to these profiles recovers both the input effective parameters, 
giving errors of $0.02\ dex$ and 0.03 mag/arcsec$^2$ in the effective diameter
and surface brightness values respectively. The total K-band magnitudes are then
estimated by extrapolating the $r^{1/4}$--law fit for individual galaxies to large
radii, using their corresponding values of the effective diameter and surface brightness. 
These magnitudes are corrected for Galactic absorption and redshift, using the
relations discussed above. Also, the effective surface brightness estimates are 
corrected for the $(1+z)^{-4}$ dimming effect. 

Finally, the K-band magnitudes, measured over an aperture of 20 arcsec.
diameter are estimated and combined with their corresponding V-band data 
from literature, 
measured over the same aperture (Lucey et al 1991). 
The $V-K$ colours were then estimated and 
corrected for Galactic extinction and redshift, using the above prescription 
for the near-ir and the relations from Faber et al (1989) for the optical
data. 
 
\subsection {The Catalogue} 

The sample of 48 elliptical galaxies in the Coma cluster, observed in this
study, is presented in Table 1. Column 2 gives the distance (in degrees)
of galaxies from the center of the cluster. In columns 3 and 4, 
the near-infrared (K-band) isophotal diameters in arcsec. 
(corresponding to a mean surface brightness of 
$\langle SB_K\rangle  = 16.5$ mag/arcsec$^2$)
and total magnitudes of ellipticals are presented. 
These are
corrected for redshift effect, surface brightness dimming and Galactic
absorption. Column 5
gives the optical V-band diameters (in arcsec.), 
corresponding to a mean surface
brightness of $\langle SB_V\rangle  = 19.8$ mag/arcsec$^2$, taken from 
Lucey et al. (1991). 
The velocity dispersion of ellipticals (log($\sigma$)- in Km/sec), 
listed in
column 6, are taken from Lucey et al. (1997). For galaxies with no available
velocity dispersion measurements from this source, the values in 
Lucey et al. (1991) were converted to the above system using the
common galaxies between the two samples. The velocity dispersions have 
an associated error of $0.03\ dex$. 
Columns 7 and 8 present the near-infrared effective diameters 
(in arcsec.) and  
effective mean surface brightnesses. This is
followed in columns 9 and 10 by the K-band aperture magnitudes and
V$-$K colours respectively, measured over an aperture of 20 arcsec. 
diameter and
corrected for Galactic extinction and redshift. The Mg$_2$ line strengths
are listed in column 11. These are also taken from Lucey et al. (1997). 
For galaxies with no such measurements from this reference, The Mg$_2$
indices from Lucey et al. (1991) are used after conversion to this system. 

\begin{table*}
\pagestyle{empty}
\caption{The near-infrared catalogue for elliptical galaxies in Coma}
\begin{tabular}{lcccccccccc}
         & r & log($D_K$)& K$_{tot}$& log($D_V$)& log($\sigma$)& log($A_e$)& 
$\langle SB_K\rangle_e$& K$_{20}$& V$-$K& 
Mg$_2$ \\
         &       &       &       &       &       &       &       &       &       &      \\
 RB45      & 0.040 & 0.810 & 11.91 & 0.858 & 2.133 & 0.859 & 16.69 & 12.31 & 3.10 & 0.280 \\ 
 N4886     & 0.050 & 0.964 & 10.86 & 1.030 & 2.209 & 1.180 & 17.25 & 11.54 & 2.95 & 0.254 \\ 
 RB43      & 0.056 & 0.799 & 12.19 & 0.866 & 2.230 & 0.694 & 16.15 & 12.52 & 3.01 & 0.262 \\ 
 N4889     & 0.060 & 1.473 &  8.20 & 1.480 & 2.595 & 1.759 & 17.49 &  9.76 & 3.32 & 0.348 \\ 
 N4874     & 0.060 & 1.266 &  8.55 & 1.294 & 2.398 & 1.934 & 18.71 & 10.44 & 3.26 & 0.323 \\ 
 N4876     & 0.062 & 0.989 & 10.89 & 1.027 & 2.267 & 1.157 & 17.17 & 11.49 & 3.19 & 0.242 \\ 
 IC4011    & 0.064 & 0.811 & 11.78 & 0.872 & 2.061 & 0.927 & 16.90 & 12.21 & 3.04 & 0.279 \\ 
 N4872     & 0.069 & 1.034 & 11.30 & 1.048 & 2.329 & 0.671 & 15.15 & 11.49 & 3.09 & 0.300 \\ 
 RB18      & 0.099 & 0.576 & 12.65 & 0.682 & 2.019 & 0.848 & 17.38 & 13.09 & 2.94 & 0.231 \\ 
 IC4021    & 0.112 & 0.944 & 11.58 & 0.943 & 2.205 & 0.711 & 15.62 & 11.85 & 3.23 & 0.299 \\ 
 N4869     & 0.119 & 1.127 & 10.27 & 1.137 & 2.303 & 1.207 & 16.80 & 11.01 & 3.23 & 0.315 \\ 
 IC4012    & 0.125 & 1.003 & 11.40 & 1.000 & 2.258 & 0.681 & 15.30 & 11.67 & 3.30 & 0.292 \\ 
 N4867     & 0.135 & 1.043 & 11.12 & 1.050 & 2.352 & 0.796 & 15.59 & 11.47 & 3.17 & 0.307 \\ 
 N4864     & 0.144 & 1.157 & 10.15 & 1.126 & 2.289 & 1.290 & 17.09 & 10.91 &  & 0.286 \\ 
 RB155     & 0.161 & 0.770 & 11.95 & 0.820 & 2.083 & 0.910 & 16.98 & 12.39 & 3.11 & 0.264 \\ 
 N4906     & 0.181 & 0.997 & 10.89 & 1.033 & 2.228 & 1.100 & 16.88 & 11.47 & 3.17 & 0.288 \\ 
 RB257     & 0.209 & 0.828 & 12.16 & 0.864 & 2.200 & 0.609 & 15.70 & 12.47 & 3.09 & 0.279 \\ 
 RB167     & 0.212 & 0.858 & 11.51 & 0.897 & 2.174 & 1.013 & 17.06 & 12.05 & 3.16 & 0.266 \\ 
 IC4051    & 0.234 & 1.107 &  9.95 & 1.087 & 2.359 & 1.448 & 17.68 & 10.99 & 3.27 & 0.333 \\ 
 D204      & 0.404 & 0.792 & 11.84 & 0.861 & 2.114 & 0.942 & 17.04 & 12.30 & 3.05 & 0.268 \\ 
 D160-100  & 0.441 & 0.945 & 11.47 & 0.959 & 2.269 & 0.786 & 15.89 & 11.82 & 3.20 & 0.285 \\ 
 n4927     & 0.464 & 1.200 & 10.16 & 1.135 & 2.450 & 1.092 & 16.11 & 10.74 & 3.50 & 0.354 \\ 
 TT41      & 0.470 & 0.725 & 12.08 & 0.790 & 2.009 & 0.919 & 17.16 & 12.51 & 3.02 & 0.260 \\ 
 D160-49A  & 0.501 & 0.914 & 11.67 & 0.961 & 2.237 & 0.742 & 15.87 & 12.00 & 3.03 & 0.270 \\ 
 N4926     & 0.565 & 1.266 &  9.80 & 1.268 & 2.420 & 1.172 & 16.15 & 10.48 & 3.32 & 0.324 \\ 
 N4840     & 0.625 & 1.169 & 10.42 & 1.166 & 2.382 & 0.983 & 15.83 & 10.87 & 3.28 & 0.319 \\ 
 D238      & 0.686 & 0.804 & 12.22 & 0.864 & 2.038 & 0.628 & 15.85 & 12.48 & 2.97 & 0.236 \\ 
 N4839     & 0.719 & 1.262 &  9.14 & 1.277 & 2.438 & 1.619 & 17.73 & 10.46 & 3.27 & 0.313 \\ 
 N4841A    & 0.723 & 1.270 &  9.53 & 1.294 & 2.414 & 1.372 & 16.88 & 10.44 & 3.24 & 0.320 \\ 
 D140      & 0.745 & 0.910 & 11.42 & 0.936 & 2.232 & 0.919 & 16.51 & 11.90 & 3.13 & 0.297 \\ 
 D160-27   & 0.768 & 0.944 & 11.30 & 0.973 & 2.235 & 0.924 & 16.41 & 11.71 & 3.14 & 0.282 \\ 
 D160-37   & 0.782 & 1.064 & 10.88 & 1.068 & 2.359 & 0.912 & 15.94 & 11.31 & 3.26 & 0.301 \\ 
 D160-23   & 0.811 & 1.003 & 11.21 & 1.011 & 2.250 & 0.865 & 16.03 & 11.58 & 3.22 & 0.310 \\ 
 N4816     & 0.839 & 1.159 &  9.95 & 1.170 & 2.330 & 1.303 & 16.96 & 10.83 & 3.24 & 0.310 \\ 
 N4824     & 0.846 & 0.912 & 11.54 & 0.929 & 2.205 & 0.830 & 16.18 & 11.83 & 3.23 & 0.278 \\ 
 IC4133    & 0.879 & 0.990 & 11.18 & 1.016 & 2.233 & 0.874 & 16.04 & 11.61 & 3.14 & 0.289 \\ 
 N4827     & 1.053 & 1.226 &  9.96 & 1.236 & 2.427 & 1.160 & 16.24 & 10.63 & 3.26 & 0.333 \\ 
 N4807     & 1.067 & 1.182 & 10.28 & 1.209 & 2.310 & 1.028 & 15.91 & 10.81 & 3.17 & 0.285 \\ 
 IC3900    & 1.171 & 1.147 & 10.51 & 1.152 & 2.428 & 0.957 & 15.78 & 10.99 & 3.23 & 0.320 \\ 
 IC843     & 1.221 & 1.238 & 10.02 & 1.170 & 2.389 & 1.100 & 16.01 & 10.61 & 3.51 & 0.303 \\ 
 N4789     & 1.525 & 1.341 &  9.30 & 1.365 & 2.416 & 1.349 & 16.53 & 10.20 & 3.25 & 0.304 \\ 
 N4971     & 1.656 & 1.080 & 10.57 & 1.072 & 2.250 & 1.096 & 16.54 & 11.21 & 3.28 & 0.291 \\ 
 D159-89   & 1.986 & 1.023 & 10.70 & 1.075 & 2.232 & 1.159 & 16.98 & 11.37 & 3.10 & 0.273 \\ 
 D159-83   & 2.500 & 1.182 & 10.23 & 1.152 & 2.306 & 1.088 & 16.16 & 10.82 & 3.38 & 0.275 \\ 
 D160-159  & 2.790 & 1.062 & 10.66 & 1.073 & 2.358 & 1.089 & 16.60 & 11.22 & 3.24 & 0.280 \\ 
 N4673     & 3.294 & 1.329 &  9.68 & 1.356 & 2.345 & 1.069 & 15.51 & 10.25 & 3.18 & 0.270 \\ 
 D159-43   & 4.545 & 1.122 & 10.60 & 1.118 & 2.399 & 0.968 & 15.93 & 11.07 & 3.33 & 0.338 \\ 
 D159-41   & 4.750 & 1.012 & 11.11 & 1.025 & 2.277 & 0.876 & 15.98 & 11.51 & 3.23 & 0.324 \\ 

\end{tabular}
\end{table*}

\section{The Near-infrared Scaling Laws}

\subsection {The Near-Infrared Fundamental Plane}

The near-infrared fundamental plane of ellipticals in
the Coma cluster is established by fitting a plane to the distribution
of galaxies on the effective diameter (A$_e$), effective mean surface 
brightness ($\langle SB_K\rangle_e$) and velocity dispersion ($\sigma$) space, 
using the data
from Table 1. A simultaneous 3-parameter fit, using 48 galaxies, 
gives
$$log(A_e) = (1.38\pm 0.26) log(\sigma) + (0.30\pm 0.02) \langle SB_K\rangle_e + c_1$$
with an rms scatter of 0.072 dex in $log(A_e)$. An edge-on view of the near-ir FP
is shown in Figure 2. 
The shape of the near-infrared FP and its scatter is in
good agreement with that found by Pahre et al (1996), using 12 
ellipticals in the Coma cluster. The presence of curvature on the infrared FP
is explored by investigating the dependence of the scatter in Figure 2 on
the FP parameters. No relation has been found between the 
residuals around the FP 
($\Delta$ (FP) $= 1.38\ log(\sigma) + 0.30 
\langle SB_K\rangle_e + c_1 - log(A_e)$) and the 
effective diameter or surface brightness. The observed rms scatter in the
near-infrared FP corresponds to an uncertainty of $18\%$ in distances to
individual galaxies from this relation. 

To investigate differences between the infrared and optical FPs, we
have also constructed the optical (V-band) FP, using the same sample of 
48 ellipticals listed in Table 1. The coefficients of the optical FP, 
estimated by performing a 3-parameter fit to these data, correspond to
1.44$\pm$ 0.04 and 0.32$\pm$ 0.01 for log($\sigma$) and 
$\langle SB_V\rangle_e$ respectively. The rms scatter in the optical FP here is
0.074 dex, which is similar to the rms scatter of 
0.08 dex (Djorgovski and Davies 1987; Jorgensen et al 1996) and 
0.07 dex (de Carvalho and Djorgovski 1992) for the B-band and 0.075 dex 
(Lucey et al 1991) for the V-band FP, using other independent samples. 

Differences between the FP coefficients from different studies, found
in literature, 
are mainly due to
the adopted fitting mathods (i.e. taking $A_e$ or $\sigma$ as the independent
variables in the least-squares solutions) or sample selection, with 
consideration to minimise the bias in the fit. 
For example, it has been estimated that a least-squares fit in $log (A_e)$, 
produces a bias of the order of 5\% for the $log(\sigma)$ coefficient, caused
by observational errors (Jorgenson et al 1996). In this study, it is attempted
to reduce such biases
by performing a 3 parameter fit to the FP. 
However, adopting
different fitting methods, the FP coefficients here
lie in the range 1.32--1.50 ($log(\sigma)$) and 0.30--0.32
($<SB_e>$) at the near-infrared and 1.23--1.44 ($log(\sigma)$) and
0.30--0.35 ($<SB_e>$) at the optical wavelengths. This shows the
sensitivity of
the FP coefficients to the adopted fitting technique, with the values
in the above range, being consistent with the present study. 
The FP parameters found here, also agree closely with other 
independent estimates
of the near-infrared (Pahre et al 1996) 
and optical (Faber et al 1987; Guzman et al 1993)
FPs.

The coefficients of the near-ir and V-band FPs are similar despite
different sensitivities to the stellar population and 
metallicity at these wavelengths. However, a smaller rms. scatter 
might have been expected in the near-ir FP since at this wavelength, the 
light samples a more uniform stellar population and is less affected by 
differences in line 
blanketing among the ellipticals. These will be addressed later in 
this section. 

Studies of the optical FP have revealed the presence of a young 
stellar population in 
ellipticals to be partly responsible for the observed scatter in
this relation (Gregg 1995b). Moreover, at a given $\sigma$, 
Coma ellipticals with lower 
Mg$_2$ values have, on average, slightly larger
$D_V$ diameters, with most of them located outside the core of the cluster
(i.e. r$>$1 deg.), indicating contributions from an intermediate age
stellar population, preferentially in ellipticals 
in low density environments (Guzm\'an et al 1992). This confirms that Mg$_2$ 
features are sensitive 
to both the `young' stellar population and local environment of ellipticals 
with the galaxies having larger contributions from the `young' stellar 
populations satisfying
$ 5.2 Mg_2 - log (D_V) - 0.442 < -0.2 $ (Guzm\'an and Lucey 1993).
There are only five galaxies in our near-infrared sample which satisfy this
relation (N4876, N4807,N4789, 159-83 and N4673) with four of them
located at the outskirts of the cluster (r$>1$ deg.). 
These galaxies are indistinguishable from the rest of the 
ellipticals on the near-ir FP in the Coma cluster (Fig 2). 

The sensitivity of the near-ir FP
to changes in the stellar population and metallicity among the ellipticals 
are further investigated by comparing the $\Delta$ (FP) values with the 
Mg$_2$ index and V-K colour 
residuals (at a given $\sigma$) from the Mg$_2-\sigma$ and
(V-K)$-\sigma$ relations respectively (Figure 3). No relation has been found. 
A similar study of the residuals diagram, using the optical FP consisting
of both field and cluster ellipticals, shows a significant trend which is
mainly interpreted as a stellar population effect (Prugniel and Simien 1996). 
However, using the sample of cluster ellipticals in this study, we find no
relation between the residuals from the V-band FP and $\Delta$ (Mg$_2$) or
$\Delta$ (V$-$K) estimates. This result has been confirmed independently
by Jorgensen et al. (1996), using a different sample of elliptical galaxies
in clusters. Therefore, it is not clear if the lack of correlation on the
residuals diagrams in Figures 3a and 3b, using the near-infrared FP in the
Coma cluster, is due to the uniformity of our cluster galaxies or, represents
a genuine effect, implying that the observed scatter on this relation is
not affected by metallicity or contributions from the young stellar 
population. These scenarios are currently explored by extending the study
of the near-infrared FP to the field ellipticals. 

However, it is likely that changes in contributions from the intermediate
age Asymptotic Giant Branch (AGB) stars to the near-infrared light of 
ellipticals is partly responsible for the observed scatter in this relation.
Spectroscopic observations of 
near-ir CO (2.3 $\mu m$) absorption features of ellipticals in clusters have 
revealed these
objects to have a smaller population of AGB stars than those in general field
(Mobasher and James 1996). Therefore, 
it may be possible to introduce near-ir CO features as an
extra parameter to reduce the scatter on the infrared FP in the same way the
Mg$_2$ indices were introduced to the optical FP to correct the relation
for contributions from the young population (Guzm\'an and Lucey 1993). 
 
Considering the measurement errors of 0.02 dex in log(A$_e$), 
0.03 dex in log($\sigma$), 0.03 mag/arcsec$^2$. in the 
$\langle SB_K\rangle_e$ and 0.02 mag. 
due to sky subtraction, and considering the fact that the errors in 
log(A$_e$) and $\langle SB_K\rangle_e$  are not independent (Kormendy 1977), 
we estimate an 
observational error of 0.040 dex for the near-ir FP. Compared with a total 
rms error of 0.072 dex, this 
gives an intrinsic scatter of 0.060 dex in the near-ir FP. 

\subsection {The $D_K-\sigma$ Relation}

The $D_K-\sigma$ relation for our sample of 48 ellipticals in the Coma cluster
is shown in Figure 4. Taking $\sigma$ as the independent variable 
(Lynden-Bell et al. 1988), a least squares fit to the $D_K-\sigma$ relation, 
using all the
48 galaxies, gives 

$$log(D_K) = (1.37\pm 0.08) log(\sigma) - (2.09\pm 0.16)$$
with an rms scatter of 0.073 dex. 
To explore the wavelength dependence of this relation, 
the optical $D_V-\sigma$ relation, using the same sample of
48 ellipticals from Table 1, is constructed. This is consistent with a
slope of $1.15\pm 0.09$ and rms scatter of 0.077 dex, in agreement
with studies based on other independent samples
(Jorgensen et al 1996; Lucey et al 1991; Djorgovski and Davies 1987). 

The similarity in the rms scatter between the two 
relations implies that 
dynamical parameters or the presence of an intermediate-age stellar
population contributing to the near-ir light of ellipticals are
responsible for the observed scatter in the $D_K-\sigma$ relation.

The slight wavelength dependence of the slope of the
isophotal diameter-velocity dispersion relation ($\sim 2\sigma$
effect) leads 
to a correlation between $log(D_V/D_K)$ and $log(\sigma)$-- (Figure 5)
as 
$$log(D_V/D_K) = (-0.18\pm 0.03) log(\sigma) + (0.42 \pm 0.05)$$

This implies that over the range of velocity dispersions covered in our
sample ($\Delta (log\sigma)\sim 0.6$), the diameters change 
by $10\%$ (ie. more massive galaxies have correspondingly
larger near-infrared diameters). Using the stellar synthesis models
of Worthey (1994), the slope of the log$(D_V/D_K)-$log$(\sigma)$ relation, due
to changes in metallicity and age, is predicted and compared with the 
observed relation in Fig 5. 
The models correspond to a range in age from 5 to 17 Gyr at a constant
metallicity of [Fe/H]=0.25 and variations in metallicity from $-0.25$ to 0.5
at a fixed age of 12 Gyr. The two models are normalised to have 
log$(D_V/D_K)=-0.01$ at 12 Gyr and [Fe/H]=0.25. These normalisations are chosen
to be close to the values expected for local ellipticals. 
The predicted relations are consistent with the
observed slope, implying that changes in age (ie. stellar population), 
metallicity or a combination of them among the ellipticals is 
responsible for the 
wavelength dependence
of the slope of their diameter-velocity dispersion relation. 
However, it is not 
possible to disentangle the effects of age and metallicity using these models. 
 
\subsection{The $L_K-\sigma$ Relation}

The total near-infrared luminosities of elliptical galaxies
in the Coma cluster are used from Table 1 to establish 
the near-ir $L_K-\sigma$ relation. The total magnitudes are employed here
in order to have this relation on the same magnitude scale as the 
previous studies of its optical counterpart.
The $L_K-\sigma$ relation is presented in Figure 6. 
A least squares fit, taking $\sigma$ as the independent variable and 
using 48 ellipticals in the Coma, gives
$$K_{tot} = (-6.78 \pm 0.50) log (\sigma) + (26.28 \pm 1.14)$$
with an rms scatter of 0.50 mag. 
Using the aperture magnitudes (measured over an aperture of 20 arcsec. diameter- column 4
of Table 1) gives a shallower slope of $-5.35\pm 0.37$ and a smaller rms scatter of 0.32 mag. 
The optical relation, using the total V magnitudes
for the same sample of ellipticals has a slope and rms scatter 
of $-6.51\pm 0.57$ and 0.49 mag. respectively. The similarity in the 
slope and rms scatter between the near-infrared and optical $L-\sigma$
relations implies that dynamical and structural characteristics in 
ellipticals are probably as important as the photometric parameters 
(i.e. stellar population, metallicity and age) in defining the mass--luminosity
relation in these systems. 
This will be explored in the following discussion. 

Taking the observational errors in K (0.03 mag.), $log(\sigma)$-(0.03 dex) and
sky background estimate (0.02 mag.), gives an observational error of 0.19 mag. 
in the $L_K-\sigma$ relation. Compared with the observed scatter 
of 0.50 mag., an intrinsic scatter of 0.46 mag. in the
$L_K-\sigma$ relation is estimated.

It is known that the colours of elliptical
galaxies are correlated with their absolute magnitudes and indirectly, with 
their mass, in both optical (Faber 1973; Visvanathan 1983) 
and infrared (Persson et al 1979; Mobasher et al 1986) wavelengths. The 
existence of such a relation has important implications for
studies of evolution of these galaxies. This colour-absolute magnitude (mass)
correlation can therefore be translated into a more direct relation between 
the colours and velocity dispersions in ellipticals. 
Using the V$-$K colours measured over an aperture of 20 arcsec. diameter for
the sample of 47
galaxies with available such data from Table 1, the $(V-K)-log(\sigma)$ 
relation is presented in Figure 7. A least squares fit gives 

$$V - K = (0.74 \pm 0.10) log(\sigma) + (1.51 \pm 0.19) $$ 
with an rms scatter of 0.083 mag. in $V-K$ colours. 
Using the stellar synthesis models of Worthey (1994), the $(V-K)-\sigma$
relations due to changes in age and metallicity are predicted and compared 
with the observed relation in Figure 7. The models are calculated for a
given metallicity, [Fe/H]=0.25, and a change in age from 5 to 17 Gyrs, 
and a range
in metallicity,$ -0.25 <$ [Fe/H] $ < 0.5$, at a constant age of 12 Gyrs. 
The models are 
normalised to have [Fe/H]=0.25 and an age of 12 Gyrs at $V-K=3.35$ mag., the
values expected for local ellipticals. They
are consistent with the observed $(V-K)-\sigma$ relation, again 
indicating that changes in either age or metallicity (or a combination of them)
among the ellipticals are, at least, partly responsible for this relation. 
However, these models cannot disentangle the effects of age and metallicity
on the $(V-K)-\sigma$ plane (Fig. 7). Recent studies also propose the presence 
of a mass-metallicity relation among the cluster ellipticals 
(Kodama \& Arimoto 1996) with the metallicity being the more 
fundamental parameter. 

To explore if the above results are model dependent, we investigate 
the behaviour of the observable parameters in ellipticals to changes in age 
and metallicity, using independent stellar synthesis models (Bruzual and
Charlot 1996). At a fixed age of 12 Gyrs and over the metallicity range
$-1.64 <$ [Fe/H] $< 1.008$, we predict a slope of 
$\Delta (V-K)/\Delta (Mg_2) = 5.23$ on the (V-K)--Mg$_2$ plane which is
close to $\Delta (V-K)/\Delta (Mg_2) = 6$ found for a fixed metallicity
of [Fe/H]=0.093 and a range in age from 5 to 12 Gyrs. 
These results are used to predict the $\Delta (V-K)/\Delta (log (\sigma))$
slope (ie. the relation in Fig. 7), using the empirical Mg$_2$-$\sigma$
relation derived from the present sample. We find that the (V$-$K)-$\sigma$ 
relations, predicted by Bruzual and Charlot (1996) model, have similar 
slopes due to changes in metallicity (at a given age) or age (at a given
metallicity). This confirms that the results in Figures 5 and 7 are 
independent of the models used (ie. the metallicity and age trends are
degenerate). 

The $(V-K)-\sigma$ relation is independent of distance and hence, 
can be used to study evolution of ellipticals at different environments
(i.e. in field and clusters) and redshifts. This relation can also
be used to estimate velocity dispersion of ellipticals at high redshifts
(for which a spectroscopic measurement of $\sigma$ is difficult), using
their V$-$K colours and allowing for evolution with redshift. 
The accuracy in log($\sigma$) values, estimated from the
(V$-$K)-$\sigma$ relation, is 0.11 dex. 

\subsection {The Near-infrared M/L Ratios}

The near-infrared FP found here differs, at a high confidence
level, from that expected from a pure virial theorem if light directly
traces the mass and if ellipticals are dynamically and structurally
homologous ($D_e\propto \sigma^2 I_e^{-1}$; where $I_e$ is the effective
surface brightness in units of luminosity per area and $D_e$ is the effective
diameter). This leads to a relation between the near-ir mass-to-luminosity 
ratios $log(M/L) = 2 log\sigma - log (D_e/2) + 0.4 \langle SB\rangle_e + a_1 $
and mass 
$log(M) = log(D_e/2) + 2 log(\sigma) + a_2$ of the ellipticals (Fig. 8)-
(Faber et al 1987; Djorgovski 1987). The $(M/L)_K$ and mass estimates in 
Figure 8 are in Solar units, assuming $z=0.023$ for the Coma and $H_0=50$ 
km/sec/Mpc. This relation is
consistent with $(M/L)_K\propto M^\alpha$ where $\alpha = 0.18\pm 0.01$, in
excellent agreement with $\alpha = 0.16\pm 0.01$ estimated from a K-band 
study of 59 ellipticals in 5 nearby clusters (Pahre et al 1996; see also
Recillas-Cruz et al 1990, 1991). 

To investigate the origin of this relation and explore if it is
caused by differences in the dark-to-luminous matter distribution
among the ellipticals or is due to changes in metallicity, age or stellar
population, we extend this study to optical wavelengths, using the same
sample of galaxies (Fig. 8). The optical relation
has a steeper slope of $\alpha = 0.23\pm 0.01$, indicating that 
over the same range in mass, the optical and infrared
M/L ratios change by factors of 3 and 2.3 respectively. 
The slight wavelength dependence of the slope of the M/L $vs.$ M relation implies that
changes in the luminous-to-dark matter distribution and/or luminosity profile
among the ellipticals is not the only parameter responsible for the observed
relations. Other observable parameters (see below), are also likely to be
as important. 

In order to have an entirely independent measurement of the slope of the 
M/L $vs.$ M relations in the two passbands, 
we estimate both the M/L ratios and masses of individual
galaxies, using their respective effective diameters in the near-ir and
optical wavelengths. However, this introduces the unphysical effect of 
the dependence 
of the dynamical mass on the wavelength in Figure 8. Therefore, the shift
in the masses of the same object between the optical and near-infrared
(M/L) $vs.$ M 
relations in Figure 8 is due to differences in the shape of the 
surface brightness
profiles in the two bands, and to observational uncertainties in the
values of the effective diameters. Nevertheless, using the 
effective diameters measured only in one passband to estimate the mass
of galaxies, the slope of the M/L $vs.$ M relations remain the same
as above, indicating that the conclusions in this section do not depend on 
different effective diameters for individual galaxies, used to estimate the mass. 

The effects of stellar population, metallicity and age on the
near-infrared mass-to-luminosity ratios of the ellipticals 
are investigated in Fig 9, 
using V-K colours and Mg$_2$
line indices. Employing the stellar synthesis models of Worthey (1994), 
changes 
in the (M/L)$_K$ with age and metallicity are calculated for 
5 Gyrs $<t<$ 17 Gyrs at constant metallicities
of [Fe/H]=-0.25, 0 and 0.25 and for a range in metallicity corresponding to 
$-0.5<[Fe/H]<0.5$ at a constant
age of 12 Gyrs. The models, normalised to have $log(M/L)_K=-0.5$
at 12 Gyrs and [Fe/H]=0, are compared
with the data in Figs 9a and 9b. 

There is only a weak dependence of 
the near-ir mass-to-luminosity ratio on metallicity. Furthermore, the
effects of age and metallicity in ellipticals, as shown in Fig. 9, decouple 
when using their (M/L)$_K$ estimates (ie. the age and metallicity sequences are
orthogonal). Assuming the same `collapse factor' (the ratio of the radius 
covering half the total mass to the effective radius) for all the elliptical
galaxies in our sample. Figures 9a and 9b suggest that the age is
responsible for the observed trends on both the (M/L)$_K-$Mg$_2$ and 
(M/L)$_K-$(V$-$K) diagrams while metallicity mainly contributes to the
scatter in these relations. 

\section{Implications of the Near-infrared Fundamental Plane}

The near-infrared FP is less affected by differences in line blanketing, 
metallicity and stellar population among the ellipticals, compared to its 
optical counterpart. This implies a FP relation with a potentially 
reduced scatter at the near-infrared wavelength. 
Therefore, the similarity of the rms scatter between the near-ir (0.072 dex) 
and optical (0.074 dex) FPs found here, was unexpected 
and could have important 
implications for constraining models of formation of ellipticals. 
Understanding the origin of this scatter provides a challenge in studying the
evolution of elliptical galaxies. 

Study of the scatter around 
the optical FP
with ellipticity (ie. shape) has found no relation between the two quantities
(Jorgensen et al 1996; Jorgensen et al 1993). 
As discussed in section 3.4, it is likely that changes in the 
$(M/L)_K$ among ellipticals is responsible for at least part
of the scatter in their near-infrared FP. This may be caused by differences 
in age and/or metallicity (Figure 9), contributions from the 
Asymptotic Giant Branch
(AGB) population to their near-infrared light or differences in matter
distribution and the internal dynamics (ie. orbital anisotropy or rotation; 
Bender et al 1993) among the ellipticals. However, similarity of the scatter 
between the optical and near-infrared FP relations indicates that structural 
and dynamical differences rather than changes in age/metallicity or the
AGB contributions among the ellipticals are probably responsible for
the observed scatter in their near-infrared FP. 

The tightness of the FP has been exploited in estimating relative 
distances between elliptical galaxies in the field or clusters. However, 
the optical FP is affected by contributions from the young population
or residual star formation. Indeed, it is likely that 
the relatively blue ellipticals, found in the Great Attractor
region, are responsible for the proposed streaming motion 
(Guzman and Lucey 1993). The near-infrared FP may be less 
sensitive to this effect (see section 3.1) and hence, the FP at this
wavelength is expected to have a more stable slope and zero-point. 
Therefore, this relation can be used to test
the reality of the streaming motion and its dependence on the non-uniformity of 
stellar populations among the ellipticals. For this purpose, we have 
performed near-ir (K-band) imaging observations of a sub-sample of 
ellipticals in the Great Attractor region. Using the present sample of 
ellipticals in the Coma cluster, we establish the zero-point of the 
field near-ir FP in order to use this to explore the sensitivity of the 
streaming motion to changes in stellar populations among the ellipticals. 
These results will be presented in a forthcoming paper.

Recent studies have indicated that elliptical galaxies are not the
homogeneous population once assumed and are likely to have a 
complex evolutionary history involving interaction, starburst
and infall of material (White and Frenk 1991; Cole et al. 1994). 
This leads to changes in the luminosity, mass and morphology of
these galaxies with time, implying that their progenitors may be
distinctly different from the galaxies we see today. A study of the FP
of ellipticals at high redshift is needed to address these scenarios. 

Recently, using the high resolution imaging capability of the 
Hubble Space Telescope and spectroscopic facilities on the largest
ground-based telescopes (MMT and Keck), 
it has become possible to explore the
evolution of the optical FP and M/L of ellipticals with redshift
(Franx et al 1996; van Dokkum \&
Franx 1996; Barger et al 1996). This reveals a FP similar to those
at low redshift with a comparable scatter. However, the observed 
change in the optical M/L is found to be smaller than that expected from
population synthesis models based on a given metallicity, IMF and 
formation redshift (van Dokkum \& Franx 1996). With the installation of
the near-infrared (NICMOS) detectors on the HST, 
it is now possible to extend this study to longer wavelengths.
The near-infrared light mainly measures the integrated
star formation and hence, is more closely related to the mass function of the
ellipticals. Therefore, study of the evolution of near-infrared FP and
M/L of ellipticals with redshift constrains the infall and merging
in these galaxies and is needed for a careful interpretation of the evolution
of the luminosity function of galaxies. 

\section {Conclusions}

The results of this study can be summarised as follows:
\begin{enumerate} 
\item the near-infrared (2.2 $\mu m$) fundamental plane of elliptical galaxies 
in the Coma cluster is studied, using a sample of 48 galaxies. This shows an 
rms scatter of 0.072 dex, similar to its optical counterpart, corresponding to
an accuracy of $18\%$ in distances derived from this relation. 

No relation is found between the scatter around the near-infrared FP and the 
V$-$K colours or Mg$_2$ line strengths in ellipticals, probably 
implying negligible
contribution to this relation due to changes in metallicity or 
the young population among the Coma cluster ellipticals. The AGB stars
are proposed as the likely source responsible for the observed scatter in
the near-ir FP. Allowing for observational uncertainties, 
we estimate an intrinsic scatter of 0.060 dex in this relation. 

\item The $D_K-\sigma$ and $L_K-\sigma$ relations are studied. These have
an rms scatter of 0.073 dex and 0.50 mag. respectively, 
compared to scatters of
0.077 dex and 0.49 mag. for their optical counterparts. 
The wavelength dependence of the slopes of these relations lead to 
correlations between $(D_V/D_K)$ and (V-K) with the mass ($\sigma$)
of ellipticals. Using stellar synthesis models, we interprete
these as due to an age or metallicity (or a combination of these two) effect. 
However, it is not possible to disentangle the effect of age and 
metallicity, using these parameters.

\item The relation between the M/L ratio and the mass of ellipticals 
($M/L\propto M^\alpha$) in the
Coma cluster is studied. The slight difference in the 
slope of this relation between the infrared ($\alpha=0.18\pm 0.01$) and 
optical ($\alpha=0.23\pm 0.01$) wavelengths is caused by
a mass-metallicity relation and not due to changes in the luminous-to-dark 
matter distribution among the ellipticals. 

\item Using the stellar synthesis models, we disentangle the effects of
age and metallicity on the (M/L)$_K-$Mg$_2$ and (M/L)$_K-$(V$-$K)
diagrams. The trend in these relations is interpreted as an age
sequence while metallicity mainly contributs to its scatter. We propose the
use of (M/L)$_K$ to separate the effects of age and metallicity in 
elliptical galaxies.

\item We propose to use the near-infrared FP of ellipticals in the Coma cluster
to fix the zero-point of its corresponding field sample in
the Great Attractor region. This is required in order to 
explore the effect of stellar population and
metallicity on the estimates of the streaming motion of galaxies. 
\end{enumerate} 

\noindent {\bf Acknowledgements:} We acknowledge support from the PPARC (BM), 
the NSF through grant AST 95-29098 (RG), 
the Royal Society (AAS) and the Space Telescope Science Institute through
a Hubble Fellowship grant HF-1055.01-93A (SZ). We are grateful to an
anonymous referee for careful reading of the manuscript.

\section{Figure Captions}
\noindent {\bf Figure 1.~(a)} 
Comparison between the K-band magnitudes of the Coma ellipticals 
in common with galaxies observed in Bower et al (1992). The magnitudes are 
measured over an aperture of $17''$ diameter, used in Bower et al.

\noindent {\bf Figure 1.~(b)} Comparison between the K-band isophotal diameters 
(at 16.5 mag/arcsec.$^2$) of a sub-set of 12 ellipticals in the Coma cluster 
observed both at the UKIRT (IRCAM3) and the AAT (IRIS).

\noindent {\bf Figure 2).} An edge-on view of the near-infrared fundamental 
plane for 48 ellipticals in the Coma cluster. The line is a plane fit to the 
data. The rms scatter in log(A$_e$) is 0.072 dex. 

\noindent {\bf Figure 3.~(a).} Deviations from the near-infrared fundamental 
plane ($\Delta (FP) = 1.38\ log(\sigma) + 0.30 <SB_K>_e + c_1 - log(A_e)$)
are plotted against $\Delta$ (V$-$K) colour residuals 
from the (V$-$K)-log($\sigma$) relation. The lack of correlation
implies that metallicity or age do not significantly contribute to the 
obseved scatter on the near-infrared FP of ellipticals in the Coma cluster. 
 
\noindent {\bf Figure 3.~(b).} The same as Fig. 3a for the 
$\Delta$ (Mg$_2$) residuals from 
the Mg$_2$--log($\sigma$) relation. Again, lack of correlation indicates that 
changes in the young population or possible environmental dependence 
among the ellipticals in the Coma cluster do not contribute to the 
scatter on their near-infrared 
FP. 

\noindent {\bf Figure 4).} The log(D$_K$)-log($\sigma$) relation for 48 
ellipticals in the 
Coma cluster. The D$_K$ diameters are measured at 16.5 mag/arcsec.$^2$ 
isophote. 

\noindent {\bf Figure 5).} The log(D$_V$/D$_K$)-log($\sigma$) relation 
for ellipticals in 
the Coma cluster. The lines are calculated from the stellar synthesis models 
of worthey (1994). They correspond to changes in age (from 5 to 12 Gyrs) at
a constant metallicity of [Fe/H]=0.25 
(dotted line) and changes in metallicity ($-0.25 <$ [Fe/H]$ < 0.5$) at a 
constant age of 12 Gyrs (solid line).
The models are transformed to the observed parameters, using the empirical 
log($\sigma$)-Mg$_2$ and log(D$_V$/D$_K$)- (V-K)
relations (from table 1), 
with the Mg$_2$ and V-K values for a given age and metallicity taken 
from Worthey (1994) models. The models are 
normalised to have log(D$_V$/D$_K$)=$-0.01$ at [Fe/H]=0.25 and 12 Gyr age.

\noindent {\bf Figure 6).} The K$_{tot}-$log($\sigma$) relation for the 
Coma cluster ellipticals. 
The total K magnitudes are used. The relation has an rms error of 0.5 mag.

\noindent {\bf Figure 7).} The (V$-$K)$-$log($\sigma$) relation for the 
Coma ellipticals. 
Colours are measured over an aperture of $20''$ diameter. 
The lines are calculated from the stellar synthesis models of Worthey (1994). 
They correspond to changes in age from 5 to 17 Gyrs at a 
constant metallicity of [Fe/H]=0.25 (dotted line) and changes in metallicity
from [Fe/H]=-0.25 to [Fe/H]=0.5 at a constant age 
of 12 Gyr (solid line).
The model estimates for $\sigma$ are calculated using the empirical 
relation as in Fig. 5. 
It is not possible to disentangle the effects of age and metallicity 
from this relation.

\noindent {\bf Figure 8).} Correlation between the near-infrared 
($\bullet$) and optical
($\ast$) mass-to-luminosity ratio; 
$log\ (M/L) = 2log\sigma - log\ (D_e/2) + 0.4\langle SB_e \rangle + a_1$
and mass; log (M) = $log(D_e/2) + 2 log(\sigma) + a_2$
for the Coma ellipticals. The same sample of galaxies are used
in both cases. 
The zero-points correspond to $a_1 = -10.67$ and $-11.24$ 
for the near-infrared and optical wavelengths respectively and $a_2=5.47$. 
Both the mass-to-luminosity ratio and mass are in solar units. 
For each wavelength, the mass corresponding to the effective diameter for that
wavelength is plotted, leading to a slight difference in the mass
estimated for the same objects in the near-infrared and optical relations. 

\noindent {\bf Figure 9~(a).} Changes in the near-infrared mass-to-luminosity 
ratios in Solar units (calculated as in Fig. 8) with 
the V$-$K colours. The lines are 
predictions from the stellar synthesis model of Worthey (1994) and correspond 
to a change in age from 5 to 17 Gyrs at a constant metallicity
of [Fe/H]=-0.25 (dashed line) and [Fe/H]=0 (dotted line)
and a change in metallicity from [Fe/H]=$-0.5$ to [Fe/H]=0.25 at a 
constant age of 12 Gyrs (solid line). 
The models are normalised to have log(M/L)$_K$=$-0.50$ at [Fe/H]=0 and
an age of 12 Gyrs. 

\noindent {\bf Figure 9~(b).} The same as Fig 9a for the Mg$_2$ 
features. The lines are 
predictions from the stellar synthesis model of Worthey (1994) and correspond 
to a change in age from 5 to 17 Gyrs at a constant metallicity
of [Fe/H]=0.25 (dashed line) and [Fe/H]=0 (dotted line) 
and a change in metallicity from [Fe/H]=$-0.5$ to [Fe/H]= 0.5 at a 
constant age of 12 Gyrs (solid line). 
The models are normalised to have log(M/L)$_K$=$-0.50$ at [Fe/H]=0 and
an age of 12 Gyrs. 
 
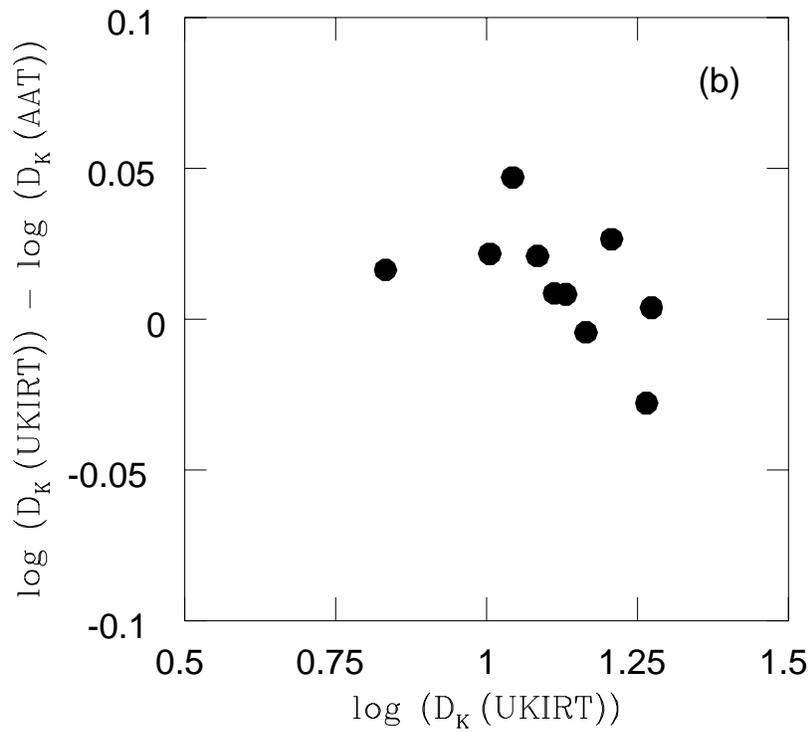
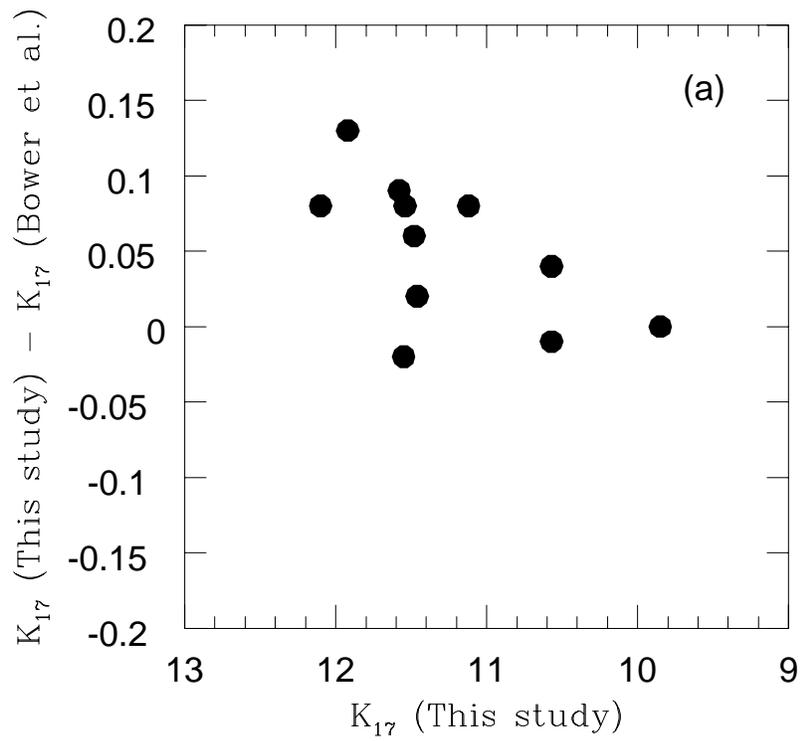
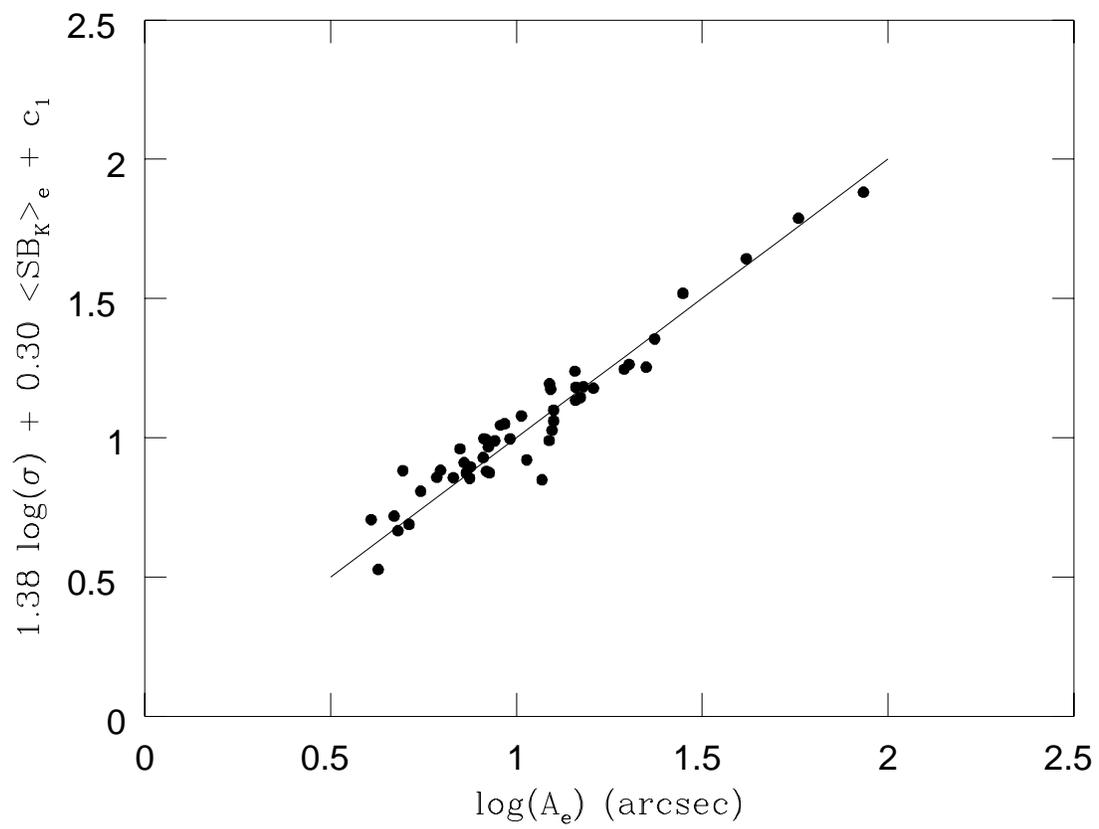
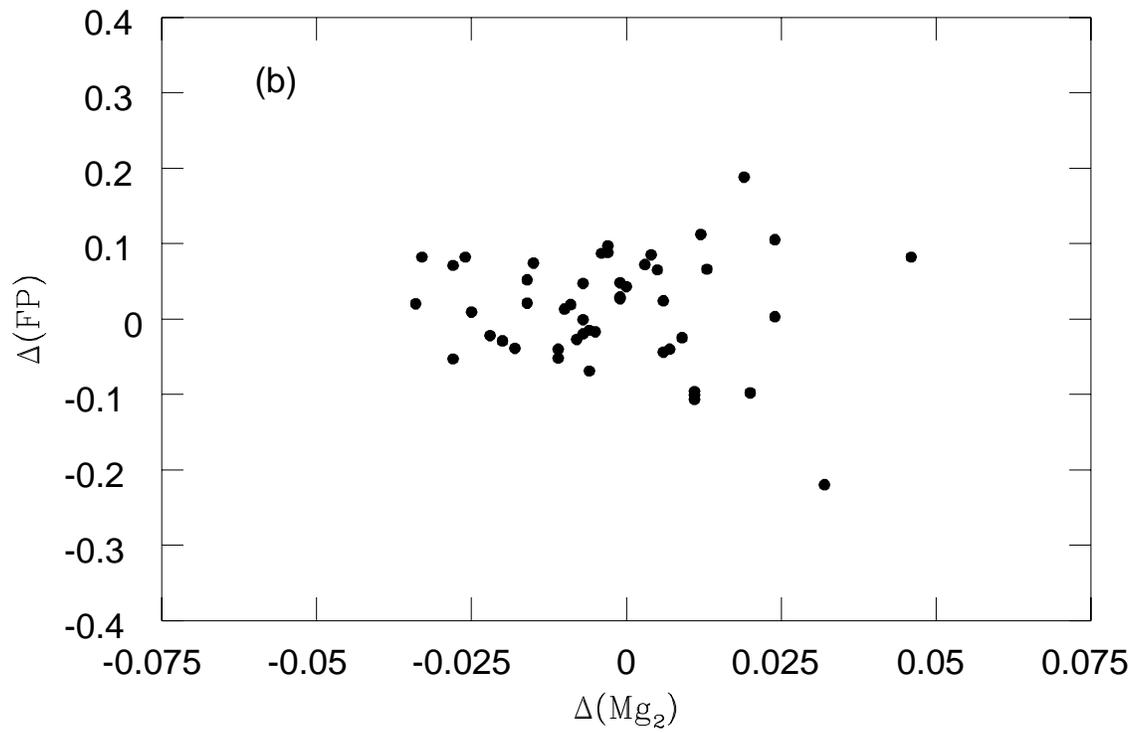
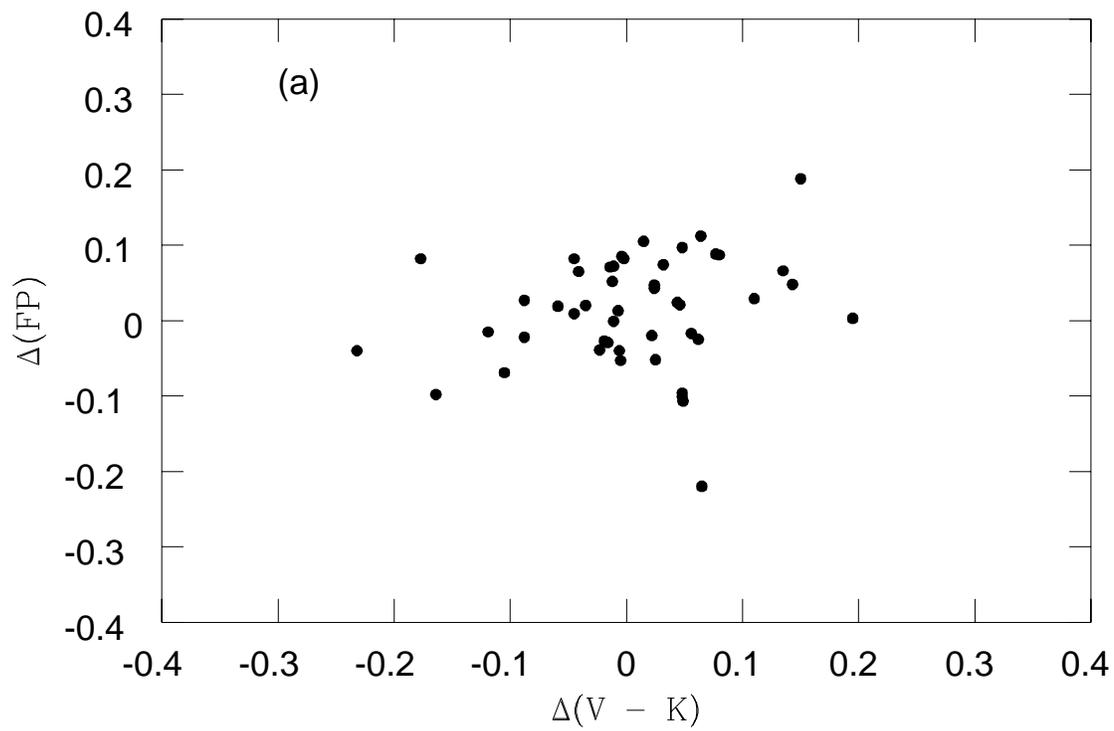
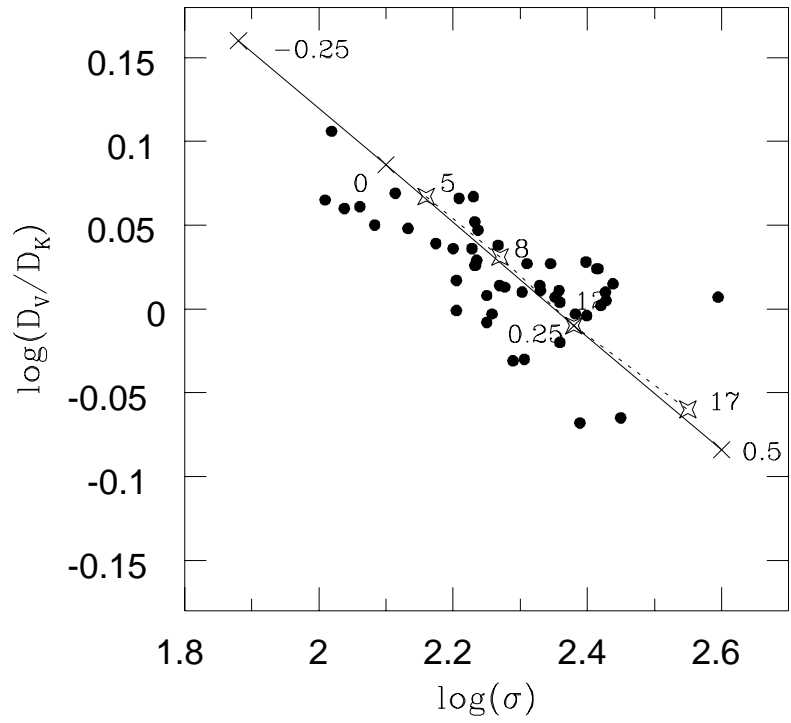
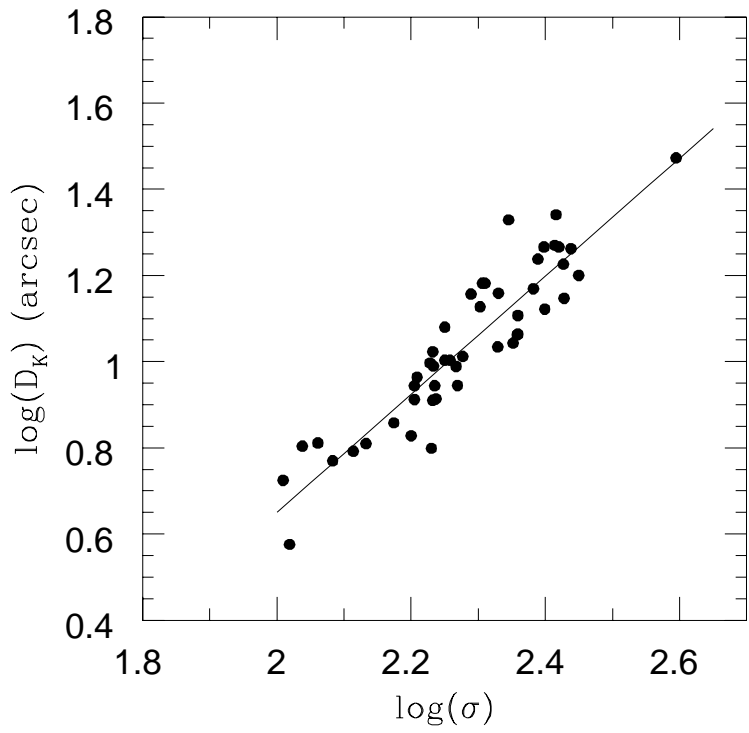
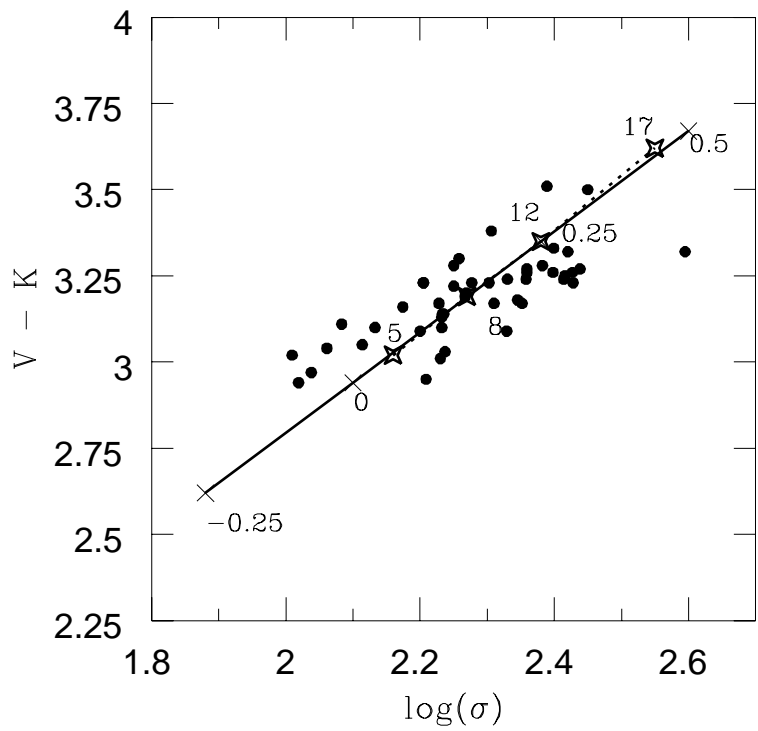
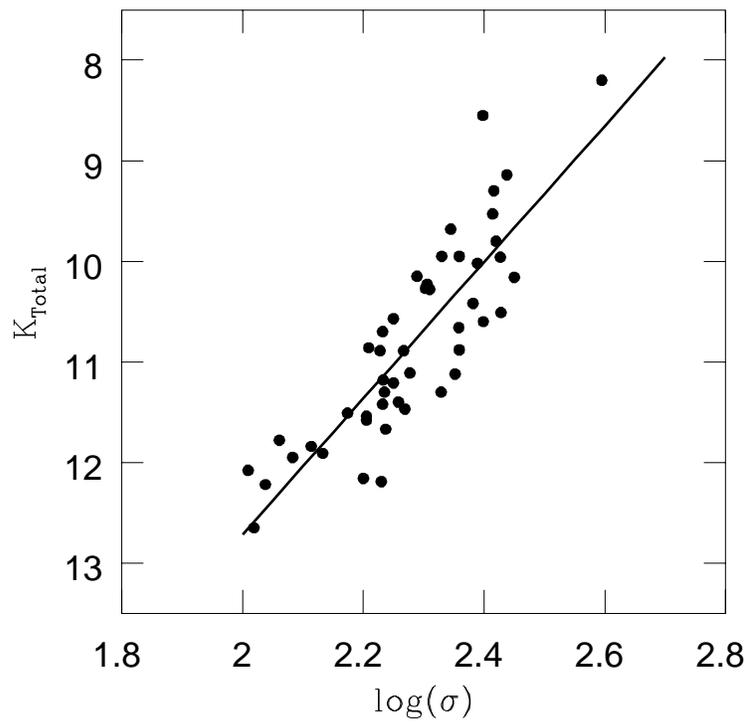
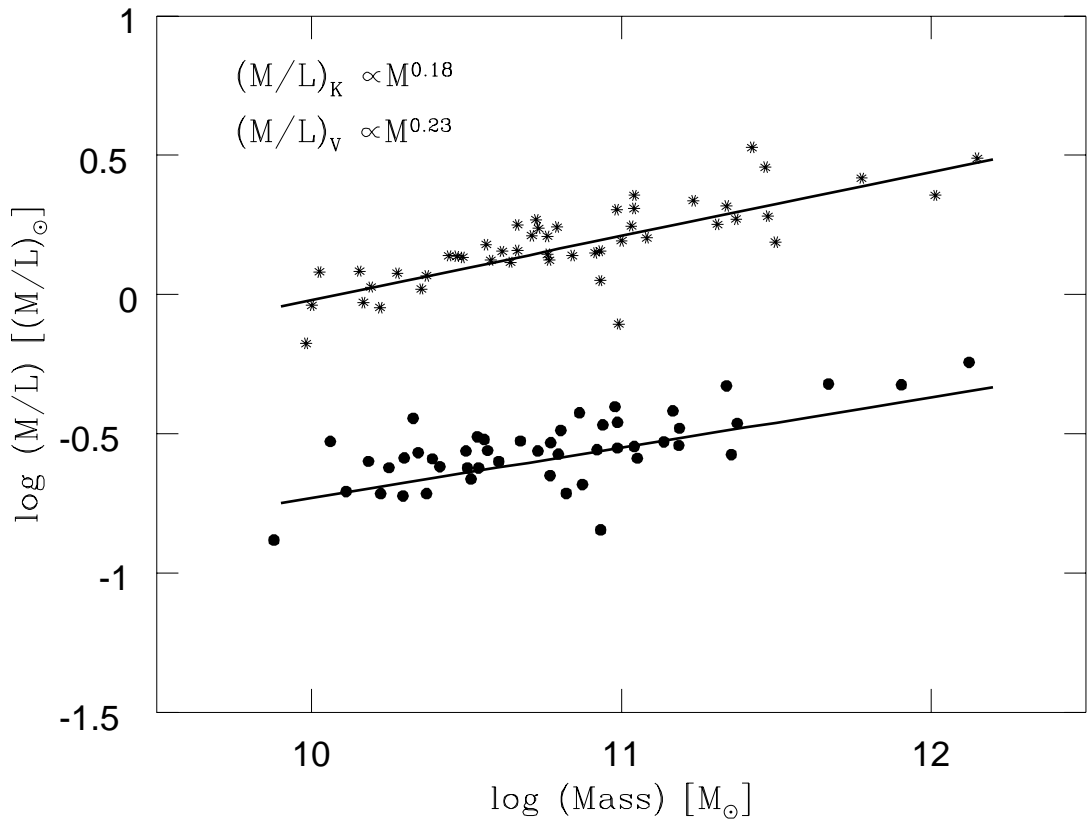
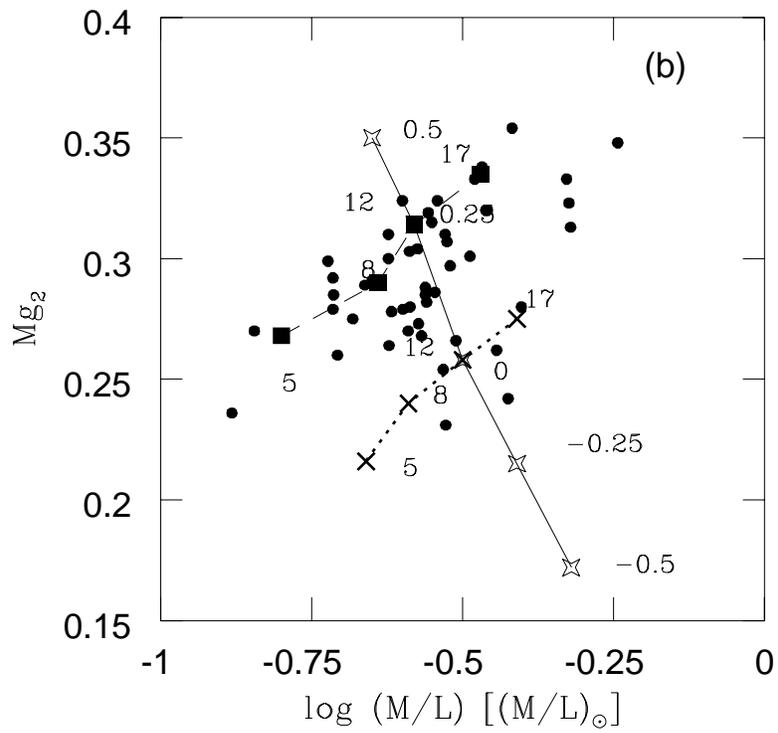
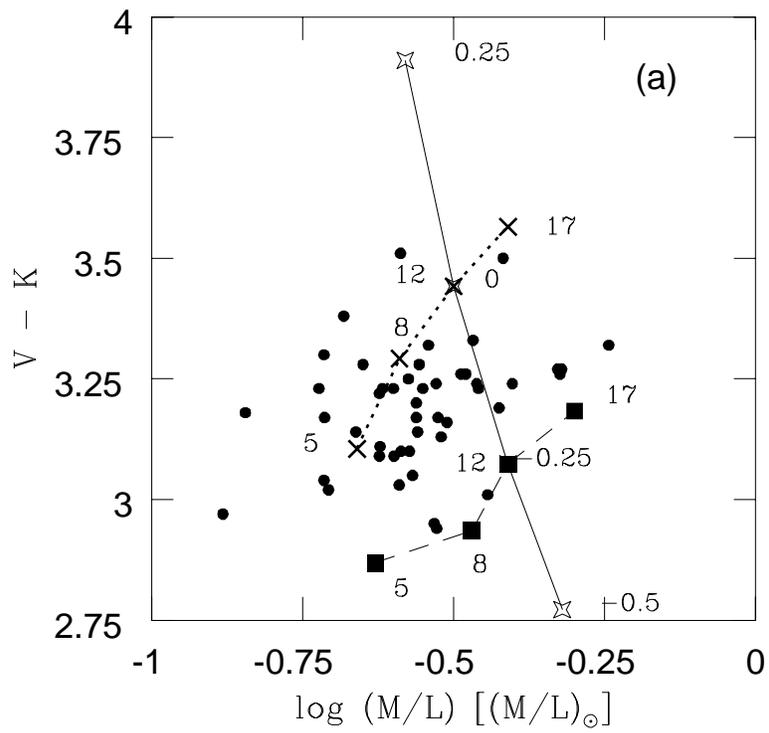
\end{document}